\documentclass[
 aip,
 rsi,
 amsmath,
 amssymb,
reprint,
]{revtex4-1}

\usepackage{graphicx}
\usepackage{dcolumn}
\usepackage{bm}

\begin{document}


\title{A scalable hardware and software control apparatus for experiments with hybrid quantum systems}

\author{Elia Perego}
\email{elia.perego@polito.it}
\affiliation{Politecnico di Torino, corso Duca degli Abruzzi 24, I-10129 Torino, Italy}
\affiliation{Istituto Nazionale di Ricerca Metrologica, Strada delle Cacce 91, I-10135 Torino, Italy}
\author{Marco Pomponio}
\affiliation{Politecnico di Torino, corso Duca degli Abruzzi 24, I-10129 Torino, Italy}
\author{Amelia Detti}
\affiliation{Universit{\`a} degli Studi di Firenze, Dipartimento di Fisica e Astronomia, via G. Sansone 1, I-50019 Sesto Fiorentino, Italy}
\affiliation{Istituto Nazionale di Ricerca Metrologica, Strada delle Cacce 91, I-10135 Torino, Italy}
\author{Lucia Duca}
\affiliation{Istituto Nazionale di Ricerca Metrologica, Strada delle Cacce 91, I-10135 Torino, Italy}
\author{Carlo Sias}
\affiliation{Istituto Nazionale di Ricerca Metrologica, Strada delle Cacce 91, I-10135 Torino, Italy}
\affiliation{LENS -- European Laboratory for Nonlinear Spectroscopy, Via N. Carrara 1, I-50019 Sesto Fiorentino, Italy}
\affiliation{INO-CNR, via N. Carrara 1, I-50019 Sesto Fiorentino, Italy}
\author{Claudio E. Calosso}
\email{c.calosso@inrim.it}
\affiliation{Istituto Nazionale di Ricerca Metrologica, Strada delle Cacce 91, I-10135 Torino, Italy}

\date{\today}

\begin{abstract}
Modern experiments with fundamental quantum systems -- like ultracold atoms, trapped ions, single photons -- are managed by a control system formed by a number of input/output electronic channels governed by a computer. 
In hybrid quantum systems, where two or more quantum systems are combined and made to interact, establishing an efficient control system is particularly challenging due to the higher complexity, especially when each single quantum system is characterized by a different timescale.
Here we present a new control apparatus specifically designed to efficiently manage hybrid quantum systems. The apparatus is formed by a network of fast communicating Field Programmable Gate Arrays (FPGAs), the action of which is administrated by a software. Both hardware and software share the same tree-like structure, which ensures a full scalability of the control apparatus.
In the hardware, a master board acts on a number of slave boards, each of which is equipped with an FPGA that locally drives analog and digital input/output channels and radiofrequency (RF) outputs up to $400$\,MHz.
The software is designed to be a general platform for managing both commercial and home-made instruments in a user-friendly and intuitive Graphical User Interface (GUI).
The architecture ensures that complex control protocols can be carried out, such as performing of concurrent commands loops by acting on different channels, the generation of multi-variable error functions and the implementation of self-optimization procedures. Although designed for managing experiments with hybrid quantum systems, in particular with atom-ion mixtures, this control apparatus can in principle be used in any experiment in atomic, molecular, and optical physics.

\end{abstract}

\maketitle


\section{Introduction}
Over the last few decades, there has been an impressive progress in the experimental study of quantum physics, specifically in our ability to control and manipulate fundamental quantum systems like atoms, ions, photons. This progress has been possible thanks to the development of several experimental techniques in Atomic, Molecular, and Optical (AMO) physics --- such as laser cooling \cite{Phillips1998}, atom and ion trapping \cite{Leibfried2003}, preparation and detection of single photons \cite{Eisaman2011}, etc. --- which are often combined together in a single experimental run. 
Experiments are performed by controlling the action of different parts of equipment (lasers, actuators, power supplies, RF synthesizers, etc.) in a synchronous sequence, in order to make ensemble measurements in which a specific temporal sequence is repeated while one or a few parameters are changed at a time.

In recent years, a great deal of development in AMO physics has grown out of research on hybrid quantum systems, i.e. composite physical systems made of two or more quantum systems realized in the same experimental apparatus. This novel approach aims at exploiting the specific features of each single system and the interactions arising between them in order to tackle physical and technological problems with new strategies. Examples of hybrid quantum systems include atom-photon interfaces \cite{Kurizki2015}, quantum gases coupled to membranes \cite{Jockel2015}, and atom-ion systems \cite{Sias2014}. Realizing experiments with hybrid quantum systems is technically very demanding, since the number of resources used in experimental setups --- such as number of lasers, RF sources, electronic equipment, etc. --- is considerably larger than in standard AMO experiments, making the realization and control of an experimental sequence more challenging.

Typically, AMO physics experiments are controlled by a single desktop computer provided with a number of modules mounted directly on the mother-board with PXI style connections or gathered in an external chassis connected to the PC. These modules can have different purposes, typically acquiring data, triggering and synchronizing devices, generating digital, analog and RF signals, and performing measurements.
The desired temporal sequence of electronic signals is typically pre-loaded into a local buffer memory, and then executed after an external trigger is provided. 

However, adapting these control systems to experiments with hybrid quantum systems is a non-trivial problem. On the one hand, the equipment that has to be used to control the experiment will in general be larger and more diverse than in ordinary AMO experiments. On the other hand, the typical timescale for the preparation of each quantum system may be very different. For instance, the typical duration of an experiment with trapped ions is on the order of tens of milliseconds, while the temporal sequence for preparing a quantum gas of neutral atoms can last up to several tens of seconds. As a result, it can be difficult to efficiently manage a hybrid quantum system with the most common control systems based on the execution of pre-loaded sequences. Consequently, the control system can become an unexpected bottleneck and a serious limitation for this kind of setup.
\\

Here we present a novel control system -- both hardware and software -- specifically designed to meet the needs of experiments with hybrid quantum systems, e.g. a mixture of trapped ions and a quantum gas of neutral atoms. 
The hardware of the control system is based on a new electronic board in which digital and analog inputs and outputs, as well as RF signals, are generated in a synchronized way. The hardware is formed by a number of identical slave boards, governed by a master board in a tree-like, scalable configuration. The boards do not only execute pre-loaded buffers, but the different input/output (I/O) channels can be independently controlled at any time. This property makes it possible to create more elaborated routines, e.g. distinct subsets of channels can execute loops in parallel, and loops can be optimized while being executed. The full action of the hardware is managed through a novel software having a base structure characterized by a tree-like structure as well. This structure ensures the possibility to manage both commercial and home-made instruments, to centralize the control and the clock signal, to perform multiple actions at the same time, and to recycle and optimize commands sequences.
\\

This paper is organized as follows.
In Sec.~\ref{sec:hardware} the hardware is presented. First, we will describe the specification of the boards and the hardware structure. Then, we will provide an insight into the main electronic board model, and present its main features.
In Sec.~\ref{sec:software} we will illustrate a novel software provided with a user-friendly GUI, and designed to be a general platform for controlling hybrid systems and AMO physics experiments.
Finally, Sec.~\ref{sec:conclusions} is devoted to the conclusions.

\section{\label{sec:hardware}
Hardware architecture
}
In general, in AMO experiments the time sequence is implemented through the synchronous action and detection of several electronic signals. These typically include analog and digital output signals -- e.g. used to drive and control the action of single actuators like piezoelectrics, or commercial devices such as synthesizers and power supplies -- and RF signals -- which can be used e.g. for spectroscopy \cite{Kasevich1989}, for manipulating atom traps \cite{Schumm2005}, or to drive acousto-optic (AOM) and electro-optic modulators (EOM).
Additionally, input and output channels are needed to acquire signals from the quantum systems, to measure and monitor the experimental sequence (e.g. by measuring magnetic fields or laser powers), and to actively control the experimental parameters through dedicated controllers like proportional-integrative-derivative (PID), or through self-optimization algorithms\cite{carleo}.

A possible strategy for realizing a hardware control system can be to assemble a custom apparatus with commercial devices.
There are several types of products available on the market, e.g. programmable generators for both digital and analog outputs\cite{DDSPulseBlaster}, or more complete control systems that offer custom solutions for pulse generation and data acquisition\cite{NIHardware, MKS, CAS, sinara}. 
These modular solutions usually come with their own software, and often with dedicated libraries for controlling them with home-made programs written in free programming languages or in commercial environments, e.g. LabVIEW (see Sec.~\ref{sec:software} for a more detailed description of possible control software solutions).
Several companies offer a broad selection of cards for acquiring data, generating digital and analog signals, triggering and synchronizing devices, with different levels of resolution and speed\cite{NIHardware, MKS, CAS, sinara}.
Nevertheless, all commercial systems have their own limitations which derive from their specific architecture.
For instance, the modular solution offered by Artiq Sinara\cite{sinara} is characterized by an architecture in which an FPGA-based master board controls a number of slave boards, each of which is tailored for a specific task (analog output, synthesizer, etc). This solution can limit the fast communication between the different input/outputs, since the data must always pass through different slots.
Moreover, many commercial solutions, like the fastest control boards of National Instruments\cite{NIHardware}, do not allow to execute concurrent tasks, which instead would be a relevant feature for controlling complex experimental setups.

Our choice is to build a control hardware as a network of FPGAs, one per slave board, in order to keep the system as flexible as possible. 
This strategy leads to clear advantages:
for instance, a fast communication between different input and output channels can be locally realized by the FPGA placed on a slave board, permitting to run simultaneous tasks, like changing the output of a channel in real time while other channels are executing other operations.

In this section we will give a general overview of our hardware architecture, and a detailed description of the boards.

\subsection{General strategy}
\begin{figure}[!t]
\includegraphics[width=0.41\textwidth]{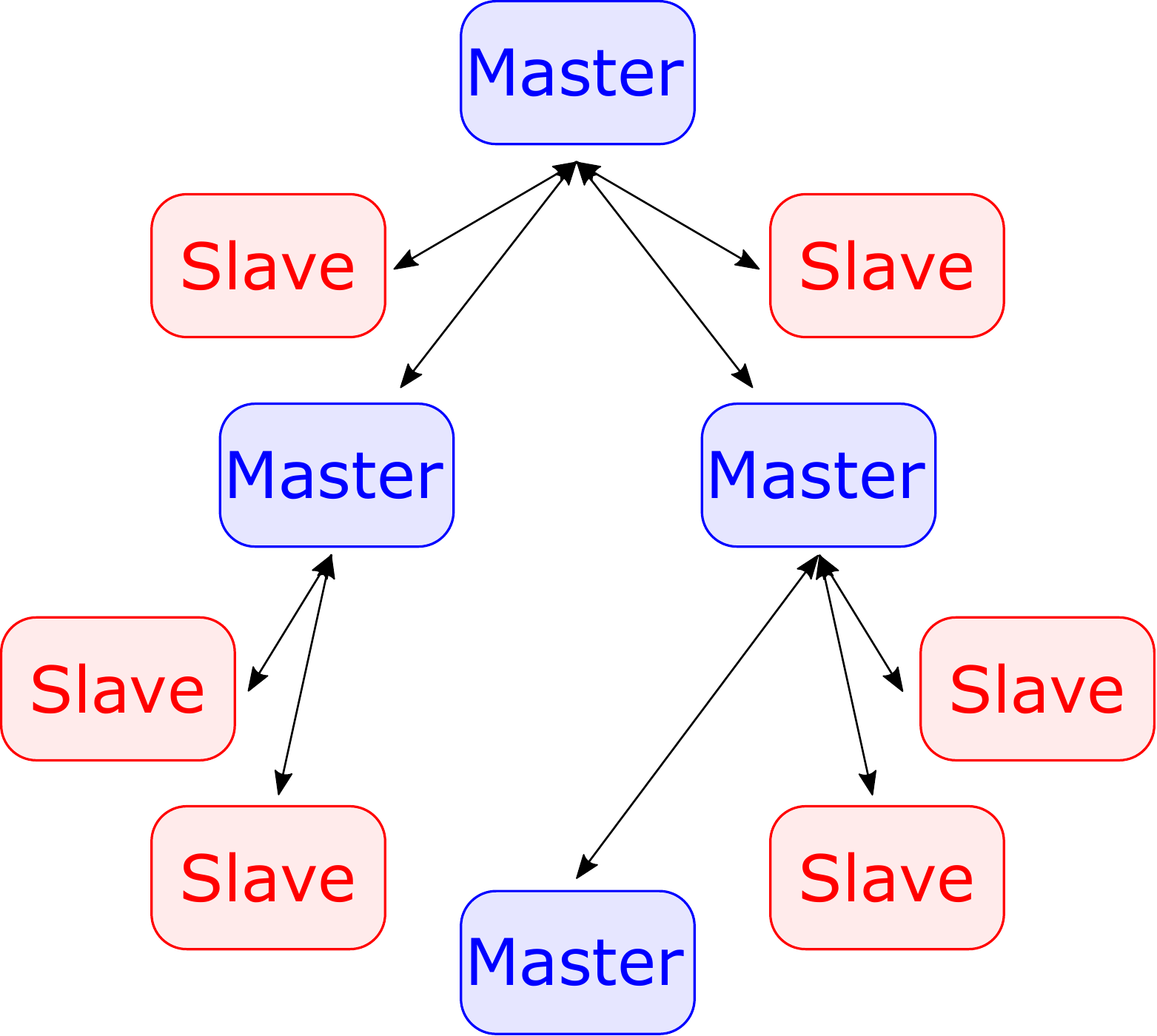}
\caption{\label{fig:TreeStructure}
Example of the tree structure of the electronic hardware. The main master board communicates to the computer and coordinates the action of up to 6 slave boards. In case more slaves are needed, additional masters can be connected to a master present at any level of the tree.
}
\end{figure}

Our strategy is to design a single electronic board (slave) that can realize -- in a limited number -- the different types of I/O signals that are needed to run an AMO experiment, and to implement it into a scalable structure so that the number of channels can be increased by adding more slots to the general hardware (see Fig.~\ref{fig:TreeStructure}).
One of the advantages of this architecture is that only two different kinds of board have to be designed, thus reducing the complexity of the electronic hardware.
The action of the slave boards is controlled and reconfigured by one or more masters, which can independently control the electronic signals generated by the slaves.
For instance, while a number of I/O channels are executing a loop, others can execute different commands in real time.

Moreover, the hardware is structured in such a way that the master (and thus the slaves) boards can be controlled also from different computers at the same time.
Finally, another relevant advantage of our hardware is that it needs a single power supply of $+12$\,V. This means that the whole system can in principle be powered by external batteries, thus making it possible to detach the system from the (noisy) line and to control portable experiments \cite{Grotti2018}.

\subsection{Hardware architecture: the tree structure}
FPGA chips are the best means to reliably produce and detect fast and precise electronic signals, since they are programmable and ensure synchrony and a high power of parallel calculus. However, driving a complex experiment with a single FPGA chip is impossible, because the finite number of I/O pins in the FPGA would limit the number of signals that can be processed. Our solution is to realize a network of FPGAs that are linked with a fast connection protocol, and that control Analog-to-Digital Converters (ADCs), Digital-to-Analog Converters (DACs), and Direct Digital Synthesizers (DDSs).

The hardware is formed by only two distinct types of modules:

\begin{description}
	\item[Master] it handles the communication with the computers by using the Local-Area Network (LAN) standard, collects the data, and coordinates the slave boards. This module is based on a System on Chip (Cyclone V SOC) that integrates an FPGA and a processor. The latter increases the versatility of the hardware, since, with respect to an FPGA, it allows to execute more complex algorithms and tasks.
	
	\item[Slave] it includes an FPGA and all the front-ends. Each module has two DDSs, four ADCs, and four DACs on board. A more detailed description of the slaves will be given below (Sec.~\ref{subseq:slaveBoard}).
\end{description}

In order to make the system scalable and the distributed computation possible, we have opted for a modular architecture of the electronics based on a tree-like structure (see Fig.~\ref{fig:TreeStructure} for an example).
The main master board is connected to the computers through a LAN and manages the entire system. Due to the limitations of the FPGA, no more than six devices (slave boards or secondary masters) can be connected to the main master.
The simplest configuration (six slaves connected to a master) is characterized by 24 ADCs, 24 DACs, and 12 DDSs used as RF generators. The number of the RF generators can be further increased by using the Numerically Controlled Oscillator embedded into the DACs, thus reaching 24 sources of RF signals.
However, if six slave boards were not sufficient, a second master can be substituted to a slave and connected to the main master; then, up to six other slaves can be linked to the second master, expanding the system to 11 slaves and 2 masters.
Interestingly, this architecture makes the system scalable without any fundamental limitation, since an arbitrary number of master nodes can be added to the tree.

Within this structure, it becomes vital that the communication between the modules is as fast as possible, in order to avoid delays and synchronization losses. To this end, we used a fast serial link (up to about 3\,Gbps) to interconnect the boards by exploiting the transceivers embedded in the FPGAs. The most efficient communication is realized between modules that are connected to the same master and physically close in space (e.g. mounted on the same rack). However, slaves on different tree levels, i.e. connected to different masters, will suffer from a reduced data bandwidth. For instance, let us imagine a certain amount of data sent from an ADC to a DAC for the purpose of creating a custom digital transfer function. In case the ADC and the DAC are on the same slave module, the full data rate (100\,MSa/s) and the lowest latency (few clock cycles) can be reached. Instead, if they are on different slave modules yet connected to the same master, the sample rate is about one order of magnitude lower, causing more latency between input and output. The bandwidth would be lowered by an additional order of magnitude if the slave modules are interconnected through different masters. Still, in most cases a ``fast" communication is needed only between elements that are physically close. Moreover, delays can be characterized and compensated for, and feedback circuits involving I/O channels on ``far" modules can easily reach 100\,kHz of bandwidth, a number that is sufficient for most applications in AMO experiments, and that can be enhanced to more than 1\,MHz when the input and output channels involved are located in the same slave board.

\subsection{\label{subseq:slaveBoard}A focus on the slave module design}

\begin{figure}[!t]
\includegraphics[width=0.47\textwidth]{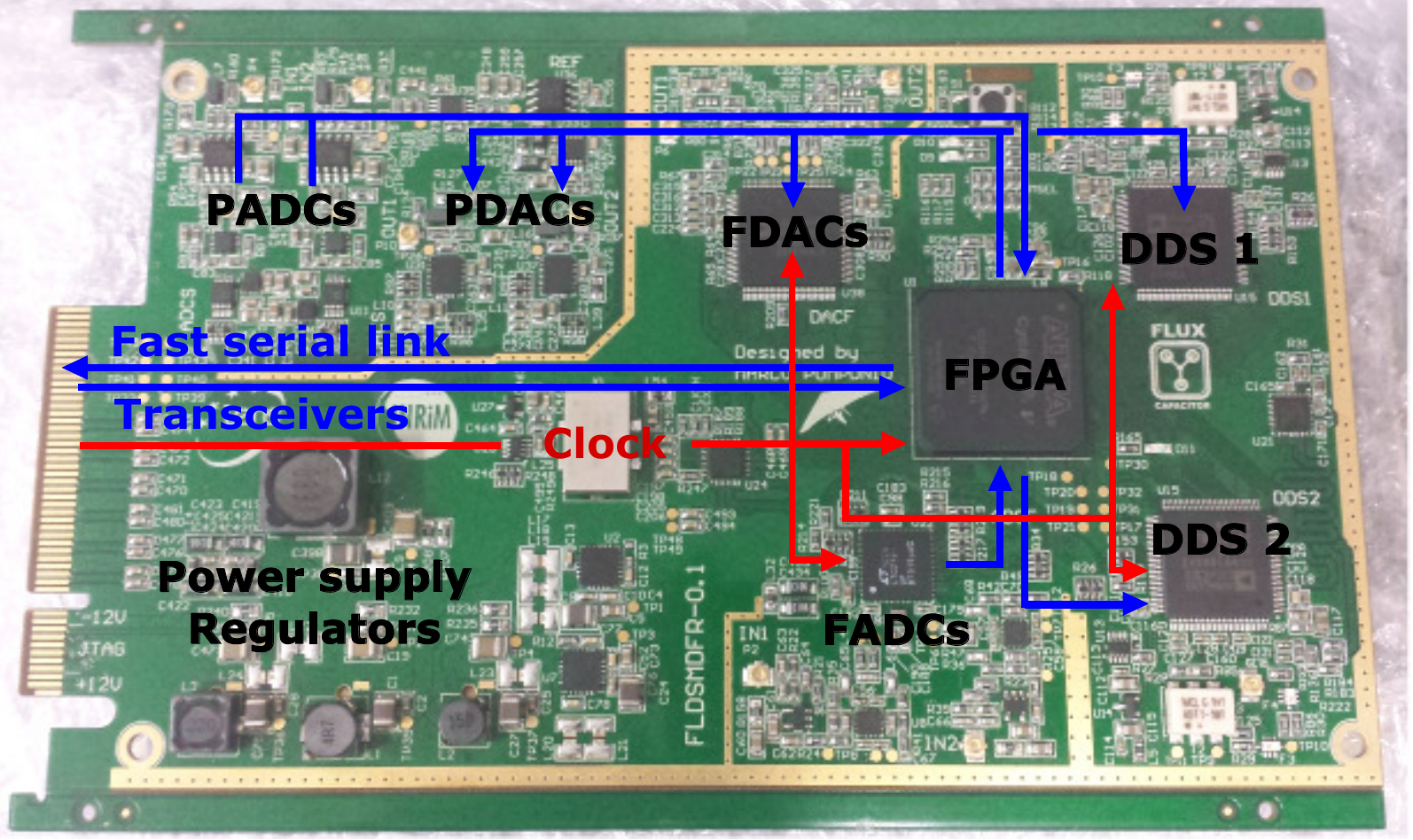}
\caption{\label{fig:SchemaScheda}
A slave module image. The board is equipped with an FPGA, two DDSs, two Precise ADCs, two Precise DACs, a Fast ADC and a Fast DAC, both of them having double inputs and outputs.
}
\end{figure}

An image of the slave module is shown in Fig.~\ref{fig:SchemaScheda}. A slave board hosts:

\begin{itemize}

\item 2 Direct Digital Synthesizers (DDS) outputs having frequency range 0.1-400\,MHz (AD9910, 1\,GSa/s, 14 bits)

\item 2 ``fast'' Analog-to-Digital Converters (FADC) with sampling frequency of 100\,MHz and resolution of 16 bits. Both FADCs are provided by a single integrated circuit (LTC2184, 100\,MSa/s, 16 bits)

\item 2 ``fast" Digital-to-Analog Converters (FDAC) outputs with sampling frequency of 800\,MHz @ x8 interpolation and resolution of 16 bits.  Both FDACs are provided by a single integrated circuit (AD9788, 800\,MSa/s, 16 bits)

\item 2 ``precise" Analog-to-Digital Converters (PADC) inputs having a sampling frequency $\sim$100\,kHz and resolution of 18 bits (AD7982, 1\,MSa/s, 18 bits, 98\,dB SNR @1\,kHz input)

\item 2 ``precise" Digital-to-Analog Converters (PDAC) outputs with sampling frequency of $\sim$100\,kHz and resolution of 18 bits (AD5780, 1\,MSa/s, 18 bits, $\pm$1\,LSB INL, 8\,nV/$\sqrt{\textrm{Hz}}$)

\item 1 FPGA communicating with the master board, governing ADCs, DACs and DDSs, and providing 2 digital inputs and 6 digital outputs (5CGXFC5C6F23C7N, 77\,k Logic Elements, 4.5\,Mb RAM, 6 transceivers, 240 I/O pins)

\item 1 ultrafast clock fanout buffer for high speed
applications (ADCLK948, 8 outputs)

\end{itemize}

We specifically devised the combination of precise but slower (18 bits, 100\,kHz sampling frequency) and fast but less precise (16 bits, 100\,MHz) ADCs and DACs on the same board in order to ensure the maximum flexibility: while for certain applications it is important to have a precise I/O and the bandwidth is non-critical or it is naturally limited by the actuator (i.e. the application of a DC voltage to a trapped ion, or the control of a magnetic field ramp), in other applications one may need to generate faster I/Os. For instance, the fast DACs could be used as RF sources for applications in which a relatively low frequency ($<$100\,MHz) is needed, e.g. for driving low-frequency AOMs, EOMs or beam deflectors, thus practically doubling the number of RF signals generated by the same slave board.
ADCs and DACs have opamp-based signal conditionings that allow to set the input/output voltage range. The latter is any voltage interval comprised within -5\,V and +5\,V.

\begin{figure}[!t]
\includegraphics[width=0.47\textwidth]{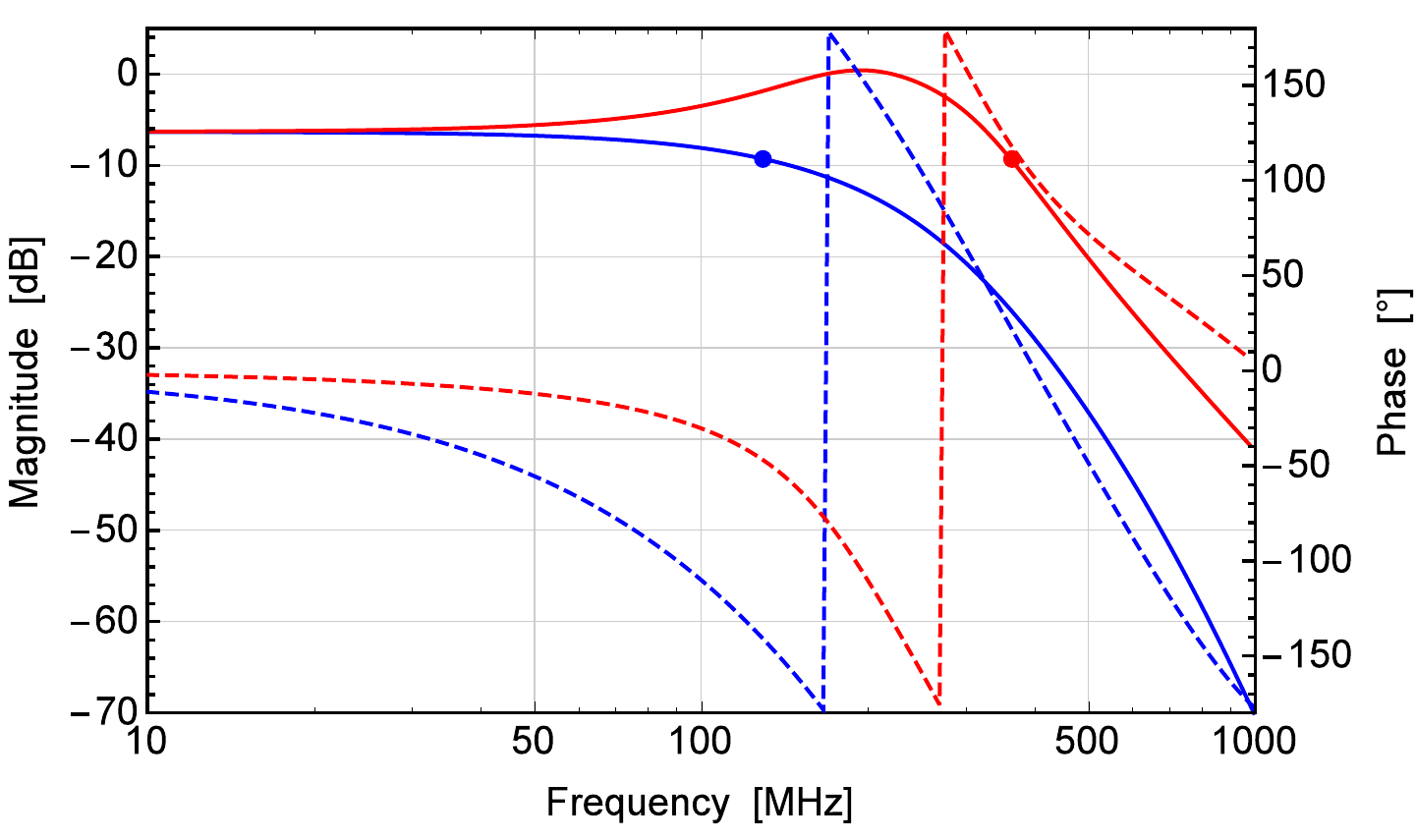}
\caption{\label{fig:FADC_bode_diagram} Tuning the FADC transfer function. In the figure we show the magnitude (solid lines) and phase (dashed lines) of the transfer functions at the minimum (130\,MHz @ -3\,dBm -- blue lines) and maximum (360\,MHz @ -3\,dBm --  red lines) achievable cut-off frequencies (blue and red filled circles, respectively).
}
\end{figure}

The front-ends of the slave board are equipped with the circuitry needed to ensure the proper operation of the corresponding integrated circuits. To this end, the DDS outputs have a transformer to convert the differential signal to a single ended one, two optional 400\,MHz low pass filters, a configurable attenuator and a fixed gain amplifier, while the FADC inputs have an opamp to uncouple the impedance and a single-ended to differential converter to provide the proper signal to the integrated circuit (LTC2184). By changing the values of capacitors and resistors, the FADC bandwidth can be tuned on the basis of specific experimental needs (see Fig.~\ref{fig:FADC_bode_diagram}).
\\

The entire system has a common reference at 100\,MHz located in the master and disseminated to the nodes of the tree-like structure with a matched length of paths (with millimeter precision). The clock distributor ADCLK948 (Analog Devices) -- present on each slot -- provides the clock directly to the FPGA, ADCs, DACs and DDSs, ensuring the synchronization of components with a time resolution of 10\,ns.
The biggest contribution to the clock skew is the latency of the components (on the order of 100\,s of ns), principally due to their pipelines. 
However, it can be compensated with the FPGA reaching values of about $\pm$5\,ns. Other minor contributions come from the skew of the components (e.g. ADCLK948 has a skew of 45\,ps), and trace lengths mismatching ($<$\,1\,mm that implies a skew less than 5\,ps). The clock jitter is less than 1\,ns.

The time resolution can be reduced to 5\,ns by using the internal phase-locked loop (PLL) of the FPGA to double the FPGA clock from 100\,MHz to 200\,MHz. 
This can be very useful e.g. to measure the fluorescence of atoms and ions with a single slave slot.
Actually, the signals of a photon counter (e.g. Hamamatsu HH11870, which provides a pulse of amplitude 2.2\,V and width 9\,ns for each detected photon, with a pulse pair resolution of 18\,ns) can be directly fed to a digital input of a slave and counted by the local FPGA, since the time resolution is smaller than the input pulses duration.
The time resolution can be further improved to the 100\,ps level by using a Vernier delay line within the FPGA\cite{vernier}. However, in such situation the delay line can suffer from possible instabilities of the power supply, which can arise if other outputs of the FPGA are concurrently used. 
In addition, the FPGA can be exploited as a counter for a limited period of time -- depending on the ``brightness'' of the source and the efficiency of the photons collection --, because of the limited size of the FPGA internal memory available for the inputs time tags collection. To circumvent this limitation, a possible strategy for increasing the time acquisition is to transfer the information from the slave board to an external memory managed by the master board onto which the data can be routed through a fast serial link.
Therefore, in experiments where the detection is realized by counting a small number of photons (up to 1 million per second) with a relatively low time resolution, our system can be extremely useful, since these data are available to perform fast operations in real-time.
For example, the FPGA counter could be used to detect the internal state of an ion with an electron shelving technique and to perform fast feedback actions on the ion following the state measurement. 
On the contrary, in experiments in which detection of the quantum system requires counting photons for a long period of time and with a high time resolution, it can be more advantageous to use an external time-to-digit converter board (e.g. TDC8HP from RoentDek Gmbh) with an external trigger option that enables a synchronization with the control system.

Another typical method for detecting particles in AMO experiments is by using CCD cameras. The operation of CCDs can be syncrhonized to the control system with the aid of external triggers, but the analysis of the data must be carried out by a dedicated software. 
\\

\begin{figure}[]
\includegraphics[width=0.47\textwidth]{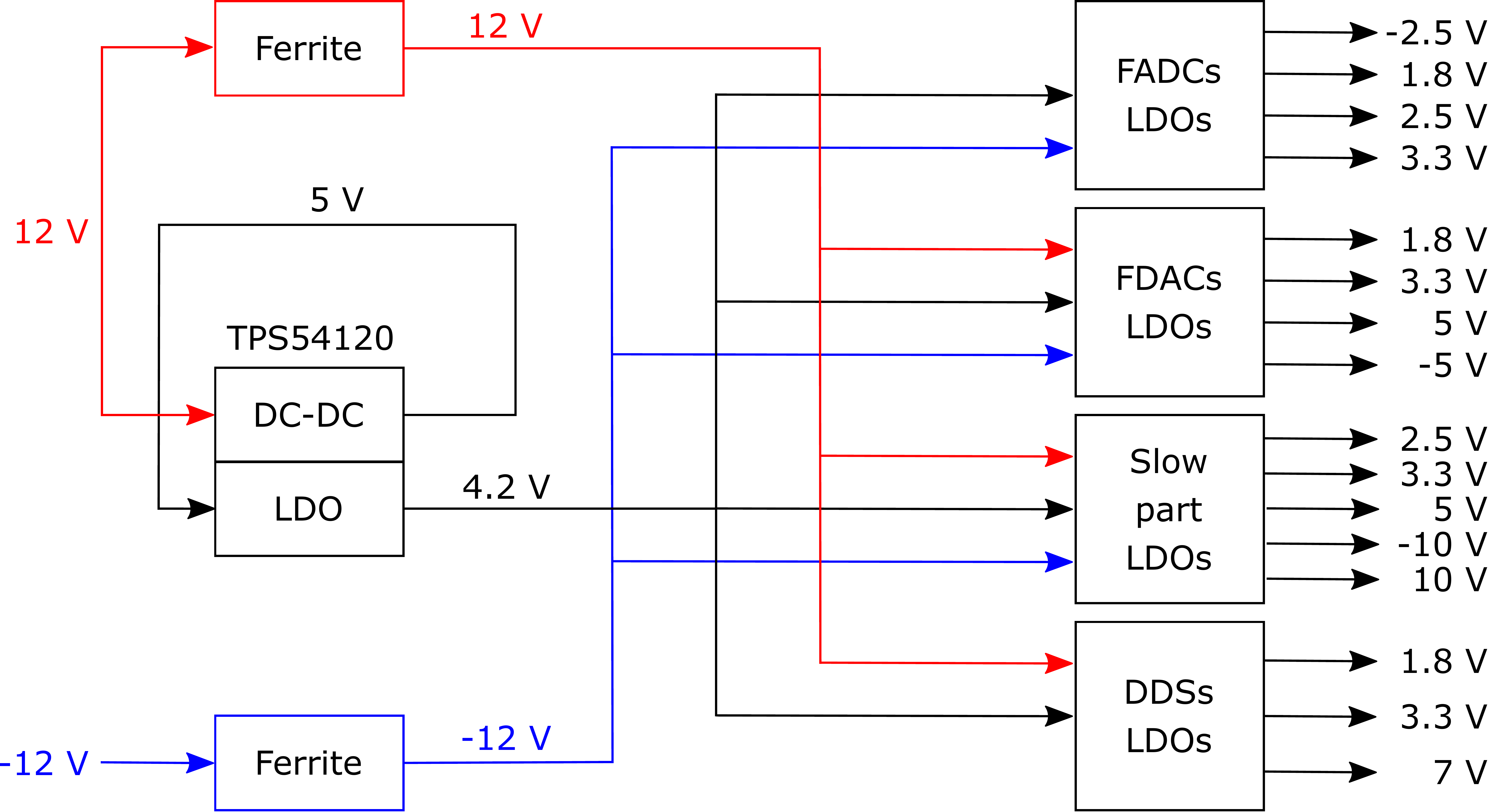}
\caption{\label{fig:AnalogNetwork} Block diagram of the analog power network implementation. The LDOs are low-drop-out voltage regulator, DC-DC are instead variable DC-DC converters, which can be regulated by the provided passive elements. TPS54120 (Texas Instruments) is a low-noise power supply that integrates DC-DC and LDO in the same component. Ferrite elements works as passive filters.
}
\end{figure}

\begin{figure}[]
\includegraphics[width=0.47\textwidth]{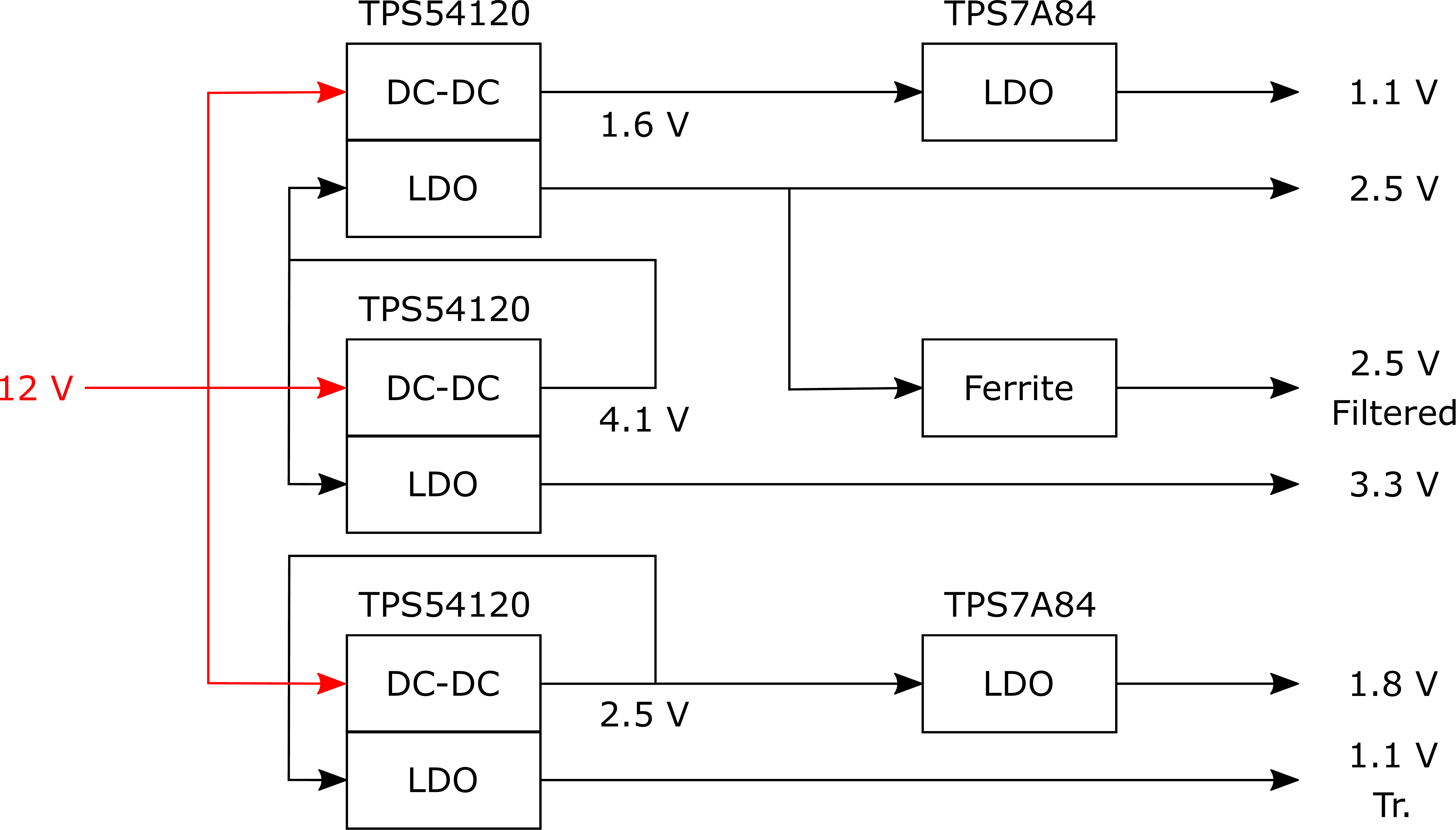}
\caption{\label{fig:DigitalNetwork} Block diagram of the digital power network implementation. TPS7A84 (Texas Instruments) is a low-noise LDO voltage regulator. For the other elements explenation, see the caption of Fig.~\ref{fig:AnalogNetwork}.
}
\end{figure}

Regarding the DDS, the AD9910 is characterized by a 14-bit DAC and a 1\,GHz internal clock. The data are loaded with 16-bit long words at a speed of 100\,MSa/s. The internal pipeline causes a data latency of approximately 100\,ns.
The output signal can be rapidly switched off by setting to zero the numerical amplitude in the dedicated DDS register, since for this value the output is null by construction.
Spurious signals related to the DDS internal operation like clock leakage can still be present, though, in general, they are not emitted at the frequency of interest. As a consequence, the RF attenuation can be considered ``infinite''. This can be very important in some applications, e.g. in avoiding light leakage from an AOM driven by the DDS.
Additionally, the outputs of all DDSs in the whole system are phase coherent, since all the slave boards share the same clock.
\\

For what concerns the power supply, the boards need to be fed only with $\pm$\,12\,V, whereas the other voltages required by the on-board components are locally generated by using a combination of switching and low-drop-out (LDO) linear regulators (see Fig.~\ref{fig:AnalogNetwork} and Fig.~\ref{fig:DigitalNetwork}). In this way, it is possible to both gain power efficiency and ensure that the components provide high resolution and low noise I/Os. In addition, in order to increase the noise performance, the analog circuitry is supplied by using different regulators from the ones used for the digital circuitry. The analog supply rails are not shared with other components, thus reducing cross-talk issues: all the required voltages are regulated close to the specific section where they are needed.
\\

In general, the PCB layout was designed to reduce digital and switching interference. For this reason, some ground plane cuts were inserted to confine the noise and protect the precision components.
The FPGA is placed at the center of the board with the fast components around, while the precision section is placed in the top left corner, protected by a ground plane cut. 
The backplane connection is operated by a PCIe$\times$8 connector. However, the module does not follow the PCIe standard, so the pin-out is custom made. This connector was chosen because it is reliable, robust, cheap, it has a large number of pins, and, most importantly, it supports fast digital signals and controlled impedance, as required by the gigabit serial interface between the slaves and the master.
Finally, the typical power consumption of each slave board is approximately 12\,W, which can increase up to 16-17\,W at full operation. In this regard, we have placed six copper layers along the whole PCB to help dissipating the heat.

\subsection{FPGA architecture}

\begin{figure}[t]
\includegraphics[width=0.47\textwidth]{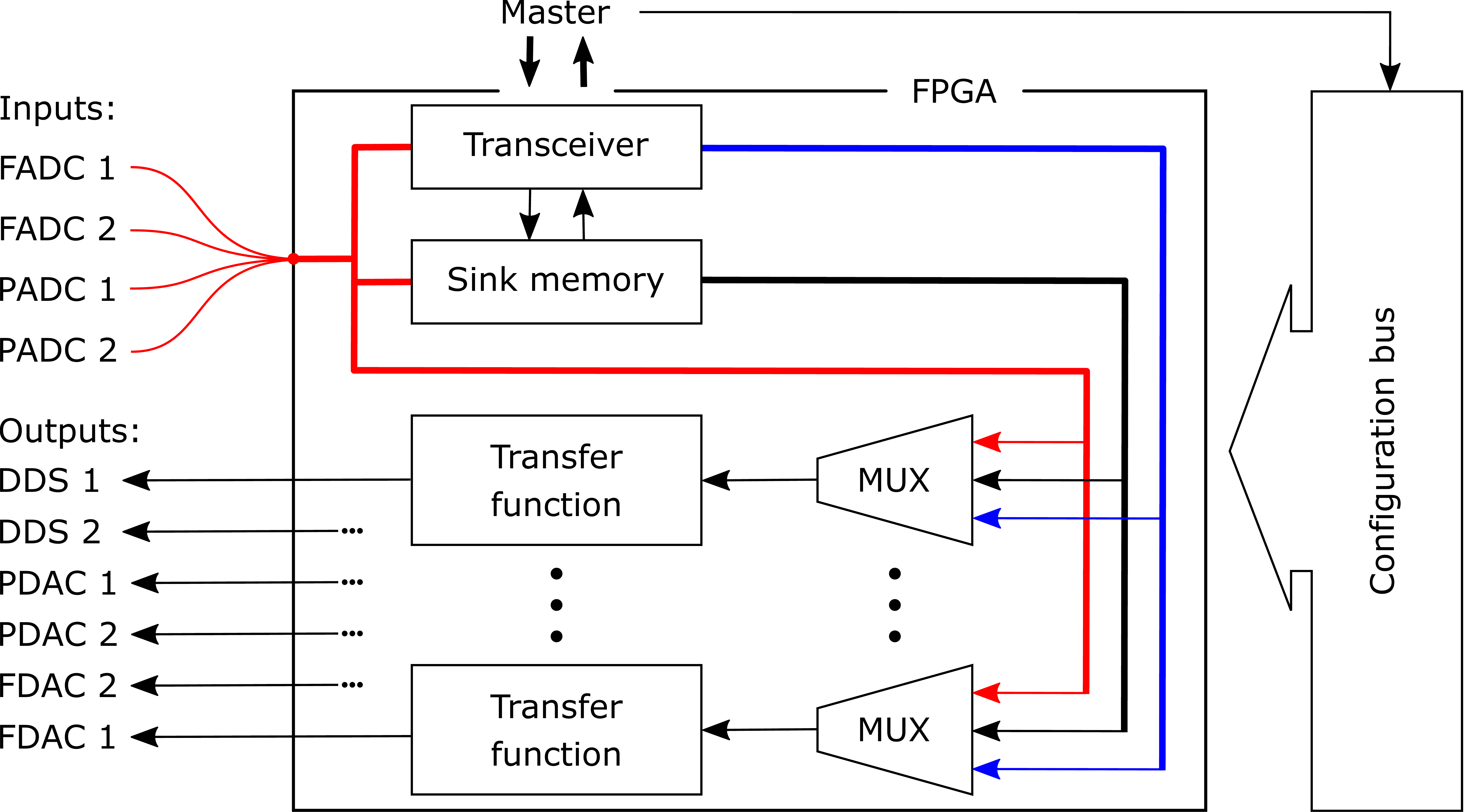}
\caption{\label{fig:FPGA_SCHEME}%
FPGA architecture. The bold lines stand for bundles of connections, whereas the thin lines represent single links.
}
\end{figure}

The FPGA configuration is composed of several logic blocks, as memories, multiplexers (MUXs), transfer functions, and drivers.
The latter are sections of the FPGA devoted to interface the FPGA with the external devices: master, ADCs, DACs, DDSs, etc. In particular, the drivers (written in VHDL) allow to adapt the input data to the component format and to translate the output of that component into a readable format, thus simplifying the incoming/outgoing communication between the units.
As shown in Fig.~\ref{fig:FPGA_SCHEME}, the output data from both the FADCs and the PADCs can be routed directly to the master through the transceiver and the fast serial link, or to a sink memory (which can be up to a tenth of the FPGA memory, since we have 4 inputs and 6 outputs), or to the DDSs and the DACs. Analogously, the output components -- DDSs, FDACs and PDACs -- can receive the input data from the input components, or from the sink memory, or from the master.
Additionally, a slower service communication, the "configuration bus", was implemented to set the parameters of the logic blocks into the FPGA.
For example, a FADC and a FDAC of a slave board can be used, respectively, to probe an experimental quantity and regulate a corresponding parameter with a PID controller. 
In this case, the data from the FADC continuously flow to the FDAC through the multiplexer and the PID transfer function, whereas their parameters (MUX address and PID gains) are set by the master through the configuration bus at the beginning of the run.
This PID regulation can be realized even with components of different slaves, but with a lower bandwidth.

Finally, the FPGA has its own embedded memory (4.5\,Mb) where the data to be generated in an experimental sequence are stored. In case the data amount does not fit the local memory, the fast serial link allows to expand the memory size by accessing the master memory.

\subsection{Slave board testing results}

\begin{figure}[]
\centering
\includegraphics[width=0.54\textwidth]{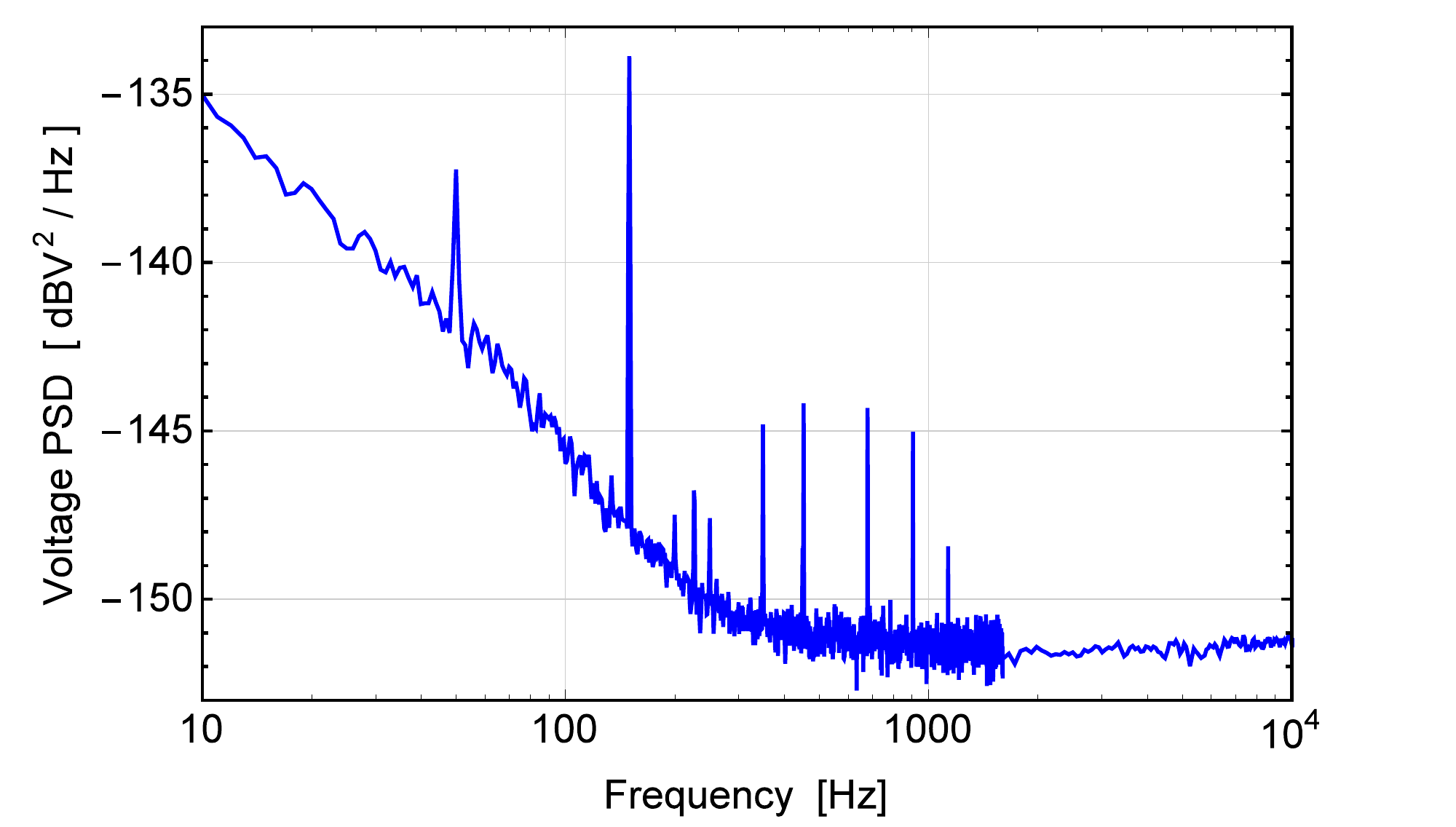}
\caption{\label{fig:DAC_noise}%
Voltage noise power spectral density of the precise DAC (PDAC).
}
\medskip
\includegraphics[width=0.54\textwidth]{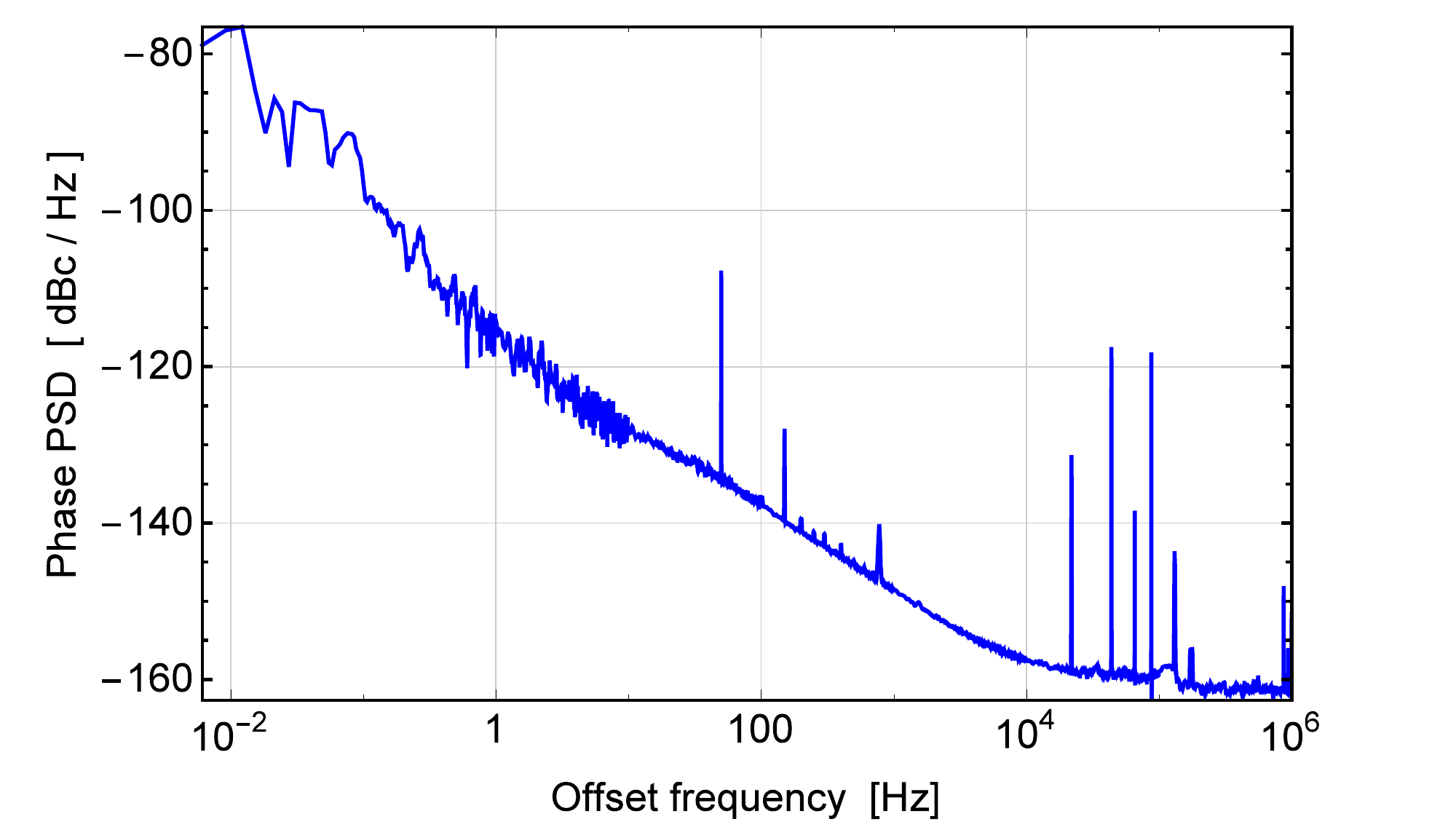}
\caption[a]{\label{fig:DDS}%
DDS phase noise power spectral density. The measurement was taken with 1\,GHz direct clock and 23.4\,MHz output frequency.
}
\end{figure}

\begin{figure}[]
\centering
\includegraphics[width=0.47\textwidth]{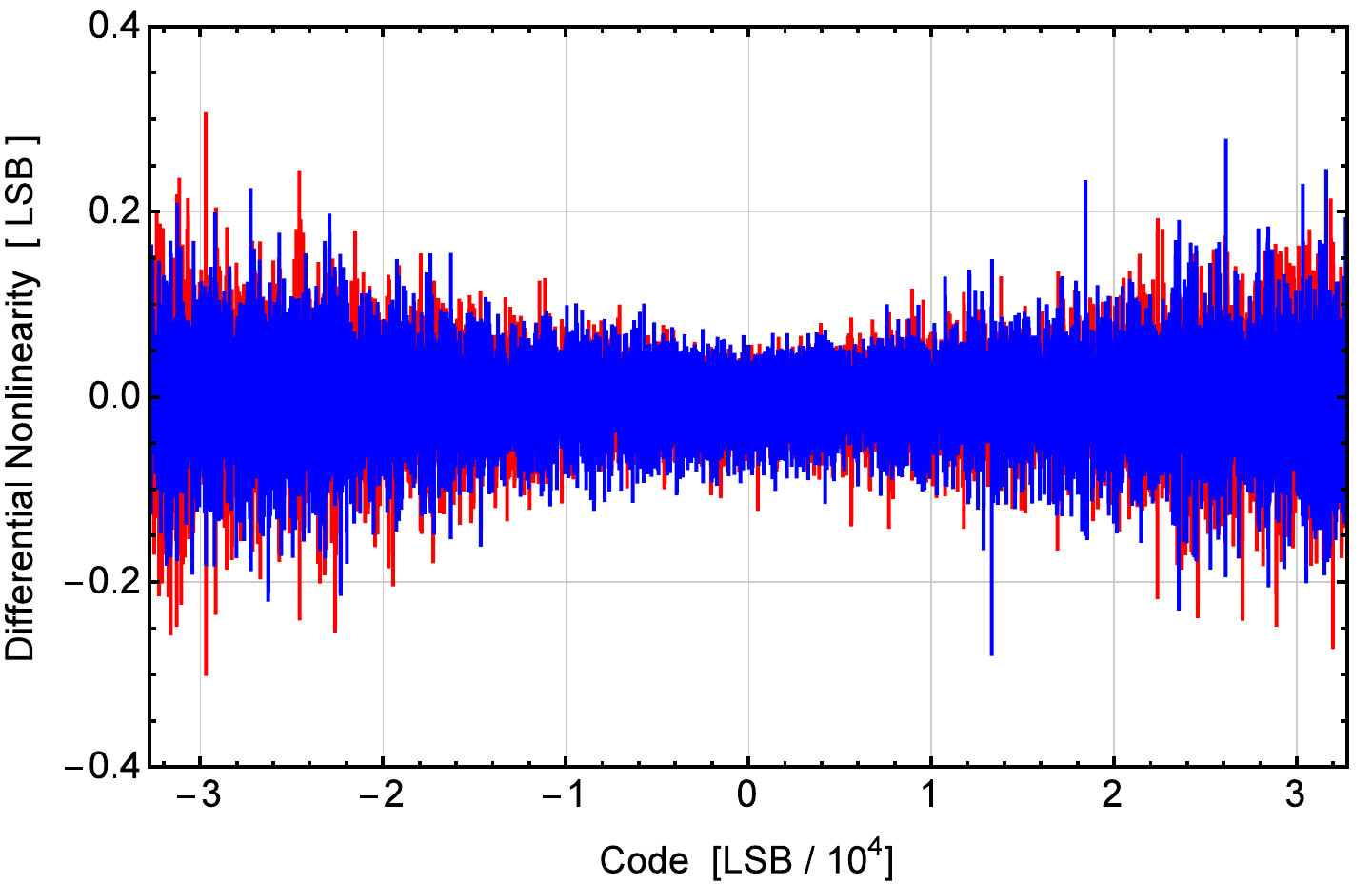}
\caption{\label{fig:DNL_FDAC}%
Differential nonlinearity as a function of the code in LSB for both FDACs present on board (channel 1/2 in red/blue).
}
\medskip
\includegraphics[width=0.47\textwidth]{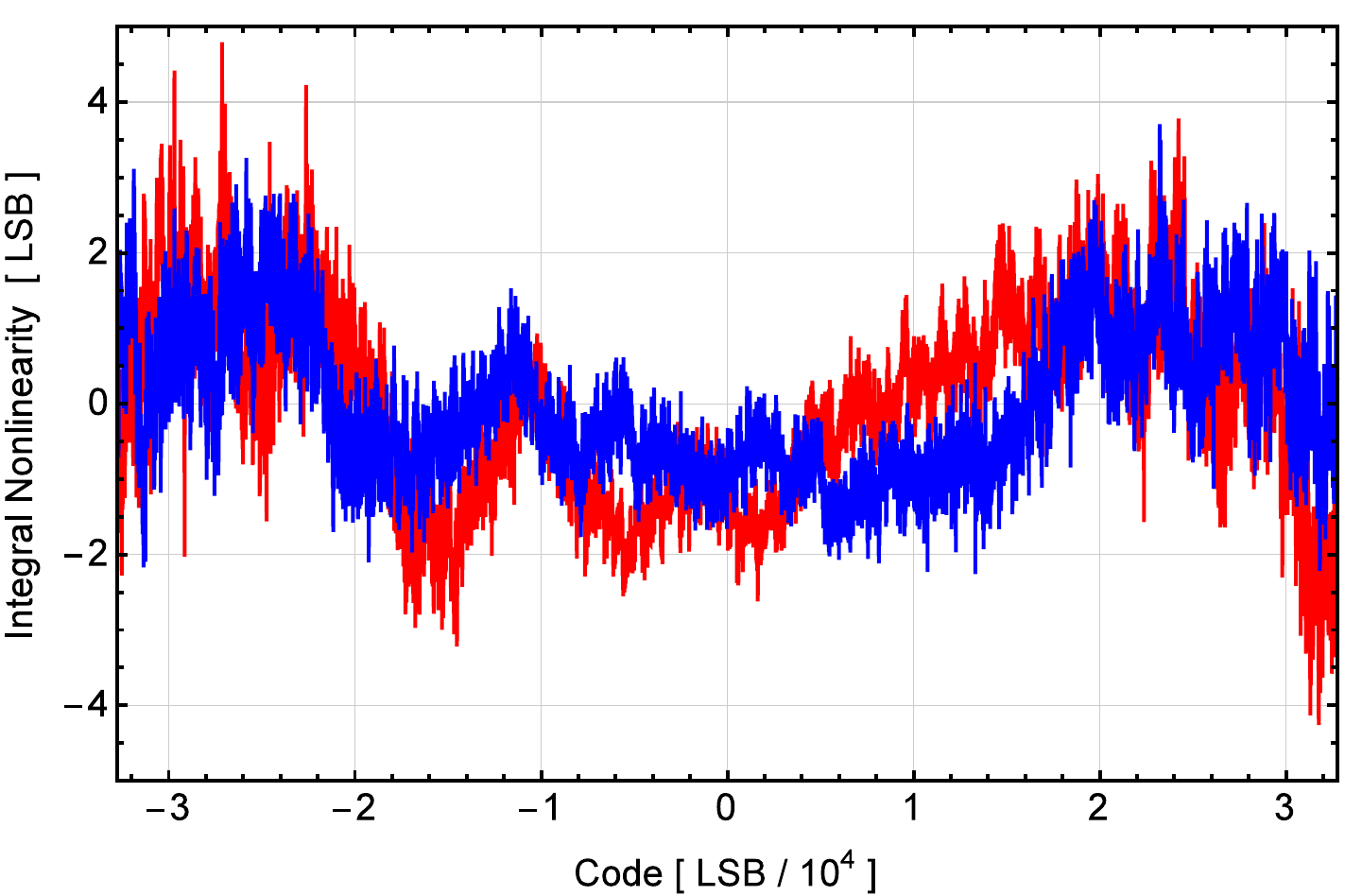}
\caption[a]{\label{fig:INL_FDAC}%
Integral nonlinearity as a function of the code in LSB for the two FDACs present on board (channel 1/2 in red/blue).
}
\end{figure}

We have characterized the performance of the I/O of the electronic hardware with several measurements. Fig.~\ref{fig:DAC_noise} shows the voltage noise Power Spectral Density (PSD) of the PDAC, while Fig.~\ref{fig:DDS} shows the phase PSD of the DDS. Both graphs reveal a low noise level and low spurs of the PDACs and DDSs outputs, as expected from the design:
there we posed particular attention in the disposition of the components on the board, and in the partition of the board in specific sectors in order to reduce the cross talk and ensure short supply rails.

Both the differential and integral nonlinearity (DNL and INL) of DACs and ADCs are within the $\pm$2\,LSB limit. As an example, Fig.~\ref{fig:DNL_FDAC} and Fig.~\ref{fig:INL_FDAC} show respectively the DNL and INL of both FDACs (800\,MSa/s). 

The latency measured between the FADC and FDAC of the same board with a direct stream of data inside the FPGA is on the order of 200\,ns. This time interval also includes the conversion from the analog to the digital domain and vice versa.

Finally, the transceivers have been tested in a loopback configuration through a 40\,cm path length using the transceiver toolkit provided by Altera. The transmission link is able to work with zero Bit Error Rate (BER) at 1\,Gbps, 2.5\,Gbps and 3.2\,Gbps, at the limit of the manufacturer specifications.

\section{\label{sec:software}
The control software: YGGDRASIL}

In typical AMO experiments, a computer is used to control the electronic hardware through a dedicated software.
To realize it, a common approach is to use commercial libraries or proprietary data-driven programming environments such as MATLAB or LabVIEW.
The latter is widely used in AMO physics to quickly implement software equipped with a graphic interface for controlling commercial boards (e.g. by National Instruments or other companies\cite{NIHardware, MKS, CAS}).
LabVIEW programs are coded in a visual language, where variables and functions are depicted as graphical objects in a canvas, arranged and connected by wires on the basis of the software aims.
The programming paradigm is based on data availability, so the sections of the program can be executed as soon as their input data are available. This inherently hinders the creation of programs where the execution of sequential operations is requested, as in AMO physics control experiments\cite{Owen2004}.
Developing scripts in MATLAB for managing instruments is also possible; however MATLAB is more frequently exploited to code programs which read data directly from the interfaced devices and elaborate them.

The alternative approach is to write the control software in a general-purpose programming language, e.g. C\#/C++\cite{Keshet2013, ControlSchreck}, Python\cite{Starkey2013, artiq} or Perl\cite{Gaskell2009}. 
Some of these options are characterized by a text-based sequence programming approach\cite{Starkey2013, Gaskell2009, artiq}, which sacrifices the graphical aspect and a user-friendly management. Moreover, some of these already existing projects are explicitly designed to control commercial devices instead of custom instrumentations\cite{Keshet2013}; whereas others do not offer enough flexible solutions, making them incapable of quickly adapting the software to hardware modifications or expansions\cite{ControlSchreck}.
\\

Given the presence of custom devices in our control system and its capability of parallel manipulation of I/O channels, we developed a new software\cite{yggdrasil} -- named ``Yggdrasil''\cite{notaPerego} -- written in C++, a widely known, well-documented and free programming language. 
The main idea behind this project is to collect in a new program some of the useful features that are already present in the similar software developed in the AMO physics community, e.g. the presence of an intuitive graphical user interface, the possibility to easily extend the software to new hardware, the ability to save optimized sequences in order to reuse them in new ones, and the possibility to implement automatized processes like self-optimization\cite{artiq}.
\\

\begin{figure}
\centering
\includegraphics[width=0.47\textwidth]{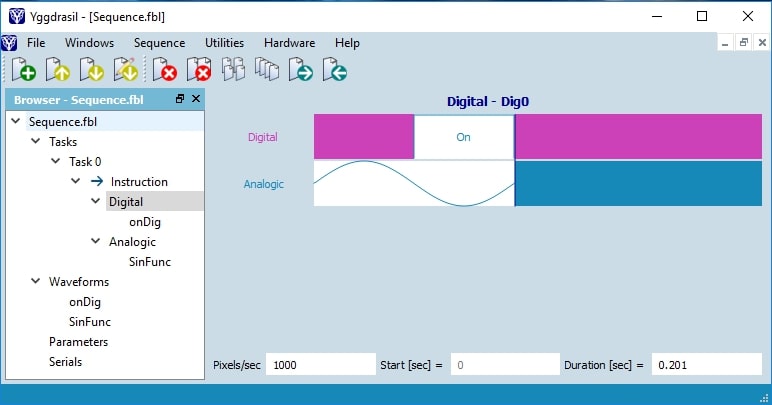}
\caption{\label{fig:ScreenShot} Screenshot of Yggdrasil's main interface. An example sequence is displayed as an open wide document. The navigation browser with the tree structure is docked on the left, while the time table representation of the instructions is on the right.}
\end{figure}

The Integrated Development Environment (IDE) used for writing, debugging, analyzing and compiling the code is Microsoft Visual Studio 2015.
For the GUI, we adopted the cross-platform application framework Qt\cite{qt} (version 5.9), for which we could use an open-source license for the purpose of developing an academic project with open-source distribution. Qt provides a number of modules containing optimized classes for managing threads, operating on data clusters and containers and creating GUIs by combining widgets. For plots and graphs representation and mathematical expressions parsing and solving, respectively ``QtCustomPlot''\cite{QCustomPlot} widget and the ``Exprtk''\cite{exprtk} toolkit library were integrated in the code.
In Fig.~\ref{fig:ScreenShot}, a screenshot of the main window of Yggdrasil is shown.

\subsection{\label{subsec:2}Yggdrasil sequence structure}

From a technical point of view, Yggdrasil GUI is based on a Multiple Document Interface (MDI) in which each independent document plays the role of a sequence. A sequence is generally defined as a set of time-ordered commands on physical channels which describe an experimental cycle. Following the Model/View/Controller (MVC) architecture pattern\cite{Reenskaug}, the backbone of a single document/sequence is a model which collects different types of data, such as digital and analog waveforms, device configurations, parameters definitions and any other detail useful to define a cycle.

Similarly to the hardware, the structure of a Yggdrasil's document/sequence has been conceived as a tree, where the root is an instance of a generalized mother class containing the necessary methods for the serialization of the data, the storage of the information and the hierarchical linking to the other nodes of the tree-like structure.
The nodes, instead, are instances of more specialized daughter classes which inherit the properties, methods and attributes of the root mother class, for the sake of code saving and efficiency.
For example, the parameters defined in a given sequence are collected in the container ``Parameters''. Within the tree representation, each parameter is a node directly linked to the common node ``Parameters''. The parameters and even their collector are all instances of the same class ``Parameter'', heir of the root class. In this way, all the parameters' data can be easily managed and retrieved, yet maintaining an essential code.
\\

\begin{figure}[]
\includegraphics[width=0.48\textwidth]{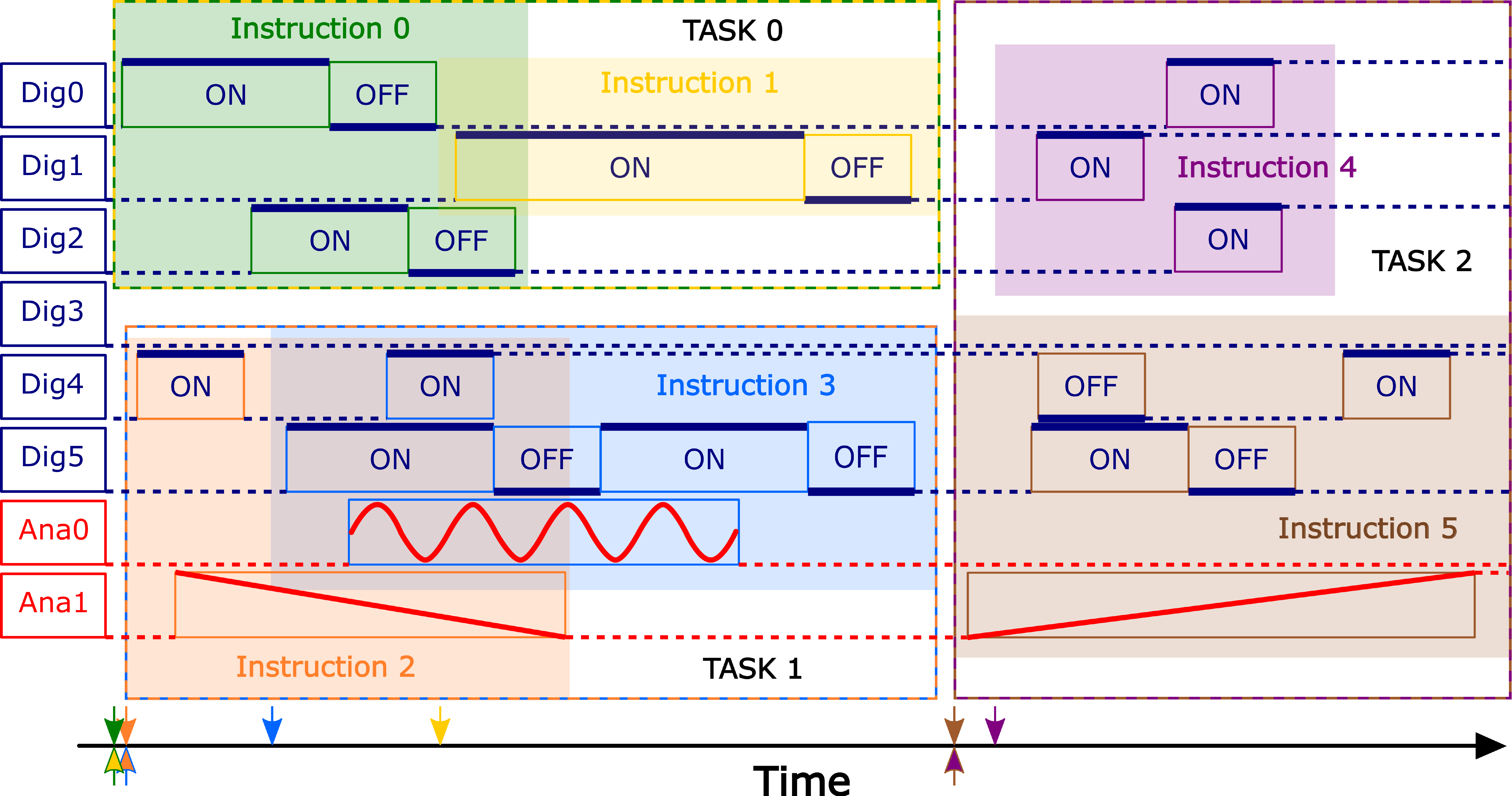}
\caption{\label{fig:SequenceBlock}%
Sketch of an artificial sequence formed by three tasks involving digital and analog tasks.
The colored arrows above/below the timeline indicate the starting points of the instructions/tasks.
Commands related to digital/analog physical channels are drawn in blue/red as lines/waveforms.
See the text for the complete description.
}
\end{figure}

In Yggdrasil, the fundamental constituents of an experimental sequence are called ``elements".
They principally contain a vector of numbers analytically or numerically defined by the user, that are executed at clock-pulse rate. The vector describes either a waveform (for an analog channel) or a composition of digital on/off pulses (for a digital channel) with a certain time duration. 
Each element can be employed by placing it in a structure named ``channel''. A channel is just a virtual container that gathers elements and empty intervals in a time-organized queue. A real physical channel is associated to the virtual channel: this operation establishes the analog or digital nature of all the data contained, and relates them to the same physical actuator.
A collection of channels can be enclosed in wider structures called ``instructions'', which are then arranged in ``tasks'', as illustrated in Fig.~\ref{fig:SequenceBlock}. The duty of an instruction is to merge each data vectors of the elements in a number of bigger vectors corresponding to the different physical channels. This merging operation is performed by respecting the ``time position'' of the elements in the channels, the time schedule of the instructions in the task and the associations between physical and virtual channels (i.e. data vectors belonging to different instructions but related to the same physical channel have to be contained in the same final data vector). In general, instructions can be overlapped in time, but it is forbidden for a channel to execute more commands at the same time.
Finally, the tasks -- which can be independently optimized to obtain a result, e.g. the creation of the largest possible quantum gas -- can be organized in a certain temporal order and in a given logic relation between each other by conditional statements. In fact, a task can represent just a section of a longer routine, for instance the loading of the atoms in a MOT or the operations which realize a quantum logic gate on a crystal of trapped ions. The temporal disposition of the tasks forms the ``sequence'', the last construct which describes the whole operations collection needed to carry out an experimental routine.
\\

An example of an artificial sequence is depicted in Fig.~\ref{fig:SequenceBlock}.
The reported sketch represents a sequence involving five digital and two analog channels.
The simple-framed rectangles in each channel stand for the elements arranged in a time order. 
The gaps between them are automatically filled with the last value of the previous element in the channel, whereas the starting default value of the channels is always zero.
As described above, by associating a physical actuator to the virtual channel (e.g. ``Dig1''), the elements (both the yellow elements and the violet one for ``Dig1'') are interpreted by the software as digital or analog commands, and executed by the same physical channel linked to the virtual one.
Elements' collections of different channels, characterized in the sketch by colors, can be encapsulated in the same instruction. The latter is a useful structure for representing more complicated commands acting on multiple channels. An example can be the production of a Coulomb cystal of ions: this operation can be performed by an instruction which contains analog signals to drive the oven current and digital ones to switch on/off the photoionization laser beam.
Finally, the instructions related to the same logic duty are gathered in a common task, represented in the sketch by the dashed-frame rectangles. For example, a task can describe all the required commands and timings to create a MOT of cold atoms or perform a quantum logic gate on trapped ions.
\\

This whole structure and all the data which define it can be serialized in raw formats and stored in a file by the software. Moreover, Yggdrasil is able to import a sequence into another one: in this way, saved and optimized tasks can be loaded as part of more complicated sequences, in which they can be combined with locally defined tasks. This makes the creation of new sequences an efficient and faster process.
\\

\subsection{\label{subsec:1}Software highligths}

The main features of Yggdrasil include:

\begin{itemize}

\item The software controls each I/O channel of the hardware, creates time sequences acting on subsets of the available channels, and makes it possible to combine loops in an intuitive and user-friendly environment. 

\item Creating automated loops of channels (both analog and digital), in which the value of a channel is automatically changed at each repetition of the loop. In this way, the software can perform a measurement without a manual action from the user.

\item Self-optimization of a specific task by performing loops in which the value of a channel is changed at each repetition, while the software monitors a measured quantity through an input channel. This feature could permit, e.g., to implement a self-optimization of a cooling cycle with the goal of maximizing the number of atoms in a quantum gas. 

\item Performing loops in parallel on different subsets of channels. This option can be useful for optimization routines in experiments with hybrid quantum systems in which there is not a common timescale for all the parts of the experiment. For instance, in an atom-ion system this feature will make it possible to use the time needed to collect atoms in a magneto-optical trap for performing independent experiments on the trapped ions.

\item Changing the value of a channel in real-time, while other channels are performing loops. With this feature a user could perform maintenance on part of the experimental apparatus, while the remaining part of the apparatus is used to perform experiments.

\item Possibility of executing loops with conditional statements based on the previous measurements. This option is used to make the software ``decide" what loop should be executed depending on the result of a measurement. In ion trapping, for instance, this option could make the system automatically run a time sequence for loading an ion in the trap if a previous measurement shows that the ion was lost from the trap. 

\item Possibility to extend the software control over equipment connected to the PC via USB or serial port, such as spatial light modulators, Arduino microcontrollers, photodiodes and commercial cameras. 

\end{itemize}

These features make the software compatible with self-learning and self-optimization processes\cite{carleo}.

\section{\label{sec:conclusions}Conclusions}
We have realized a new control system -- composed by both an electronic hardware and a control software -- designed for managing experiments with hybrid quantum systems.

The hardware is composed of two main types of boards, a slave board that realize all the I/O channels, and a master board that interconnects and governs the slave boards. The I/O channels include low noise analog and digital signals, and RF outputs. The boards can be arranged in a hierarchical structure that ensures scalability and a fast and efficient operation of the channels thanks to the on-board FPGAs.

The software shares the same tree-like structure of the hardware, and is specifically designed to exploit all the features of the control electronics. These features include the possibility of changing in real-time the value of a channel while other channels are performing a loop, the possibility of combine loops of channel subsets in order to perform concurrent experiments, and the possibility of performing self-optimizing routines that automatize part of the experimental work. 
Although the control system was conceived for controlling a hybrid system of ultracold atoms and ions, it can be in principle used to control any AMO physics or transportable experiments.

We hope that the ideas here described can be useful and inspiring for other experimentalists that desire to build their own control system.
\\

\begin{acknowledgments}
We thank M. Inguscio for continuous support, the members of the LENS electronic workshop for discussions, and E. Bertacco for his invaluable help. This work was financially supported by the ERC Starting Grant PlusOne (Grant Agreement No. 639242), the project EMPIR 17FUN07 (CC4C), the SIR-MIUR grant ULTRACOLDPLUS (Grant No. RBSI14GNS2), and the FARE-MIUR grant UltraCrystals (Grant No. R165JHRWR3). This project has received funding from the EMPIR programme co-financed by the Participating States and from the European Union's Horizon 2020 research and innovation programme.
\end{acknowledgments}

\bibliography{Yggdrasil_bib}

\end{document}